# Controllable Water Vapor Assisted Chemical Vapor Transport Synthesis of WS$_2$-MoS$_2$ Heterostructure

*Yuzhou Zhao and Song Jin\**

Department of Chemistry, University of Wisconsin – Madison, 1101 University Avenue, Madison, Wisconsin 53706, United States

*Corresponding authors. E-mail: jin@chem.wisc.edu

**Abstract:** The vapor phase synthesis of two-dimensional transition-metal dichalcogenides (MX$_2$) and their heterostructures is often poorly reproducible and sensitive to uncontrolled environmental humidity. It was recently realized that water vapor can play important roles in the growth of MX$_2$ by reacting with MX$_2$ at high temperature to form volatile metal oxyhydroxide (MO$_x$(OH)$_y$) and hydrogen chalcogenides (H$_2$X) that dramatically change the growth processes. Here we report the controllable synthesis of WS$_2$, MoS$_2$, and their heterostructures using water-assisted chemical vapor transport (CVT). The water vapor can be tunably delivered by thermal dehydration of calcium sulfate dihydrate (CaSO$_4$·2H$_2$O) solid precursor, which not only provides much lower vapor pressure baseline and a wider tunable range than liquid water, but also can be readily integrated into a chemical vapor deposition process. This allows controlled growth of monolayer, multilayers, and spiral nanoplates of $WS_2$, as well as the lateral epitaxial growth on the edge of $MX_2$, and more reproducible growth of large area $WS_2$-$MoS_2$ heterostructures. Raman and photoluminescence spectral mappings confirm the various types of $WS_2$-$MoS_2$ heterostructures. These results reveal insights into the growth mechanisms of $MX_2$ and provide a general approach to the controllable growth of other metal chalcogenides.

Two-dimensional (2D) transition metal dichalcogenides (TMDs, generally expressed as $MX_2$, M =Mo, W, *etc.* ; X = S, Se, Te) and their heterostructures have attracted increasing research interests in recent years due to abundant choice of materials,[1-3] their fascinating electro-optical properties,[1, 4-8] and potential applications in high-performance and flexible optoelectronic devices.[1, 4-8] While mechanical exfoliation of high-quality single crystals[9-11] is commonly used for fundamental studies and liquid exfoliation[12-13] has been demonstrated for mass production, vapor phase growth provides scalable approaches to grow high-quality and large-area 2D single crystals with atomic layer thickness.[3, 14-18] The conventional vapor phase growth methods for $MX_2$ can be categorized by the precursors used in reactions, such as metal oxides,[16-27] metal chlorides,[28-30] metal-organic precursors,[31-34] as well as metal dichalcogenides.[20, 35-41] Chemical vapor deposition (CVD) synthesis of $MX_2$[15, 42] have been commonly used to successfully enable diverse morphologies, such as monolayers,[16-18, 25-26] monolayer polycrystalline thin films,[33-34, 43-44] spiral nanoplates,[39-40, 45-48] and lateral[19-22, 35-37] and vertical[21-22, 28-29] heterostructures, *etc*.

Despite these successes, the vapor growth of 2D $MX_2$ layers and their heterostructures are notoriously difficult to reproduce. Water vapor as a common but fluctuating atmospheric constituent could be one of the many causes of the irreproducibility in vapor phase growth of $MX_2$. It is believed that most of the precursors used for $MX_2$ synthesis are also sensitive to water vapor and environmental humidity during the reaction, where the presence of water reacts with the precursors causing hydrolysis or oxidations.[28, 35, 49] Uncontrolled moisture could be the reason that causes $MX_2$ vapor phase growth less reliable and reproducible. So researchers are always trying to minimize the effect of moisture, often by keeping the reactor under vacuum for long time or baking the reactor before reactions. Although the importance of minimizing moisture is commonly accepted in the community, the role water plays in the reactions is less studied. It was only until recently that the moisture level was carefully monitored during synthesis or water vapor was intentionally used in the growth of MX$_2$ and their heterostructures.[28, 35, 49] For example, the correlation between the ambient humidity and the effective partial pressure of the MoCl$_5$ precursor during the reaction was reported, and layer-controlled MoS$_2$ vertical heterostructure can be synthesized by carefully tracking the humidity and compensating its effect with reduction in the reaction time or precursor temperature.[28] Furthermore, it was reported that water vapor can react with MX$_2$ at high temperature to form volatile metal oxyhydroxide [MO$_x$(OH)$_y$] species and hydrogen chalcogenides (H$_2$X) that dramatically change the growth mechanisms and increase the effective vapor pressure of the growth precursors.[35] By utilizing water vapor generated from a water bubbler and hydrogen gas controlled by mass flow controller, exquisite multijunctions of lateral heterostructures were recently realized.[35] Interestingly, water vapor has also been previously identified as a crucial parameter in the CVD synthesis of single-walled carbon nanotube,[50-52] graphene,[53] and hexagonal boron nitride,[54] where it reacts as a gentile oxidizer and etching regent.

On the other hand, many recent studies report physical vapor deposition (PVD) of MX$_2$ using MX$_2$ powder as the precursor in ambient pressure CVD setups.[20, 36-41] However, considering the high melting points of MX$_2$ (MoS$_2$: 2380 °C; WS$_2$: 1800 °C),[55] and the high melting temperatures of the corresponding metal element (Mo: 2623 °C; W: 3422 °C), the reported reaction temperatures seem rather low for a PVD mechanism in ambient pressure. In fact, the thermal stability of MX$_2$ in ambient or vacuum has been controversial.[55-56] It is possible that such confusing thermal stability could be caused by the trace water vapor present in those experiments. This also suggests that some growth reactions that were believed to be simple thermal PVD could actually go through a chemical vapor transport (CVT) pathway with water vapor actually serving as the transport regent.[20, 36-41] Such water-assisted CVT hypothesis also suggests the importance of not only minimizing the water vapor before or during reactions but also controlling water vapor intentionally to control the synthesis and improve the reproducibility.

In this work, we report the controllable synthesis of $WS_2$ monolayers, few layers, and spiral nanoplates, as well as various types of $WS_2$-$MoS_2$ heterostructures using water-assisted chemical vapor transport. We utilize the thermal dehydration of solid water vapor source calcium sulfate dihydrate ($CaSO_4 \cdot 2H_2O$) to conveniently control the amount of water vapor. This facile approach not only provides much lower vapor pressure baseline and a wider tunable range than liquid water precursor, but also can be easily integrated into a CVD process without further modifying the setup. This allows the precise control of the growth of various $WS_2$ (and other $MX_2$) monolayer and nanostructures, the lateral epitaxial growth of $MoS_2$ on the edge of $WS_2$, and more reproducible growth of 3 different types of large-area heterostructures of $WS_2$-$MoS_2$. Raman and photoluminescence (PL) mappings confirm the different types of heterostructures. Our research provides chemical insights into the growth mechanism of $MX_2$ materials and it can potentially be a general approach to controllable grow other $MX_2$ materials.

We generate water vapor from $CaSO_4 \cdot 2H_2O$ solid, also known as gypsum in its mineral form. When gypsum is heated, the water of hydration is released to produce $CaSO_4$, as shown below (the partially dehydrated gypsum is also known as plaster of Paris):

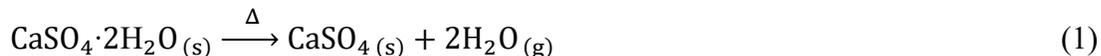

$$CaSO_4 \cdot 2H_2O_{(s)} \xrightarrow{\Delta} CaSO_{4\,(s)} + 2H_2O_{(g)} \qquad (1)$$

The anhydrous $CaSO_4$ is powerful in absorbing water at room temperature thus can be effectively used as a common desiccant commercially known as "Drierite". Conversely, it is difficult to dehydrate $CaSO_4 \cdot 2H_2O$ at room temperature, however, when $CaSO_4 \cdot 2H_2O$ is heated, it can be used to generate water vapor with different rates at different temperatures. Here we first examined the ability of $CaSO_4 \cdot 2H_2O$ to release water vapor at different temperature. Figure 1a shows the percent mass change when 10 mg $CaSO_4 \cdot 2H_2O$ was heated and monitored as a function of time at different isothermal temperatures under constant $N_2$ flow of 100 sccm in a thermogravimetric (TGA) instrument. Relatively constant dehydration rate can be observed at each temperature from 80 to 120 °C, which shows that heated $CaSO_4 \cdot 2H_2O$ can serve as a stable water source during the reaction. The weight percent stabilized at 79.1%, consistent with a final dehydrated product of $CaSO_4$ as shown in the Equation 1.

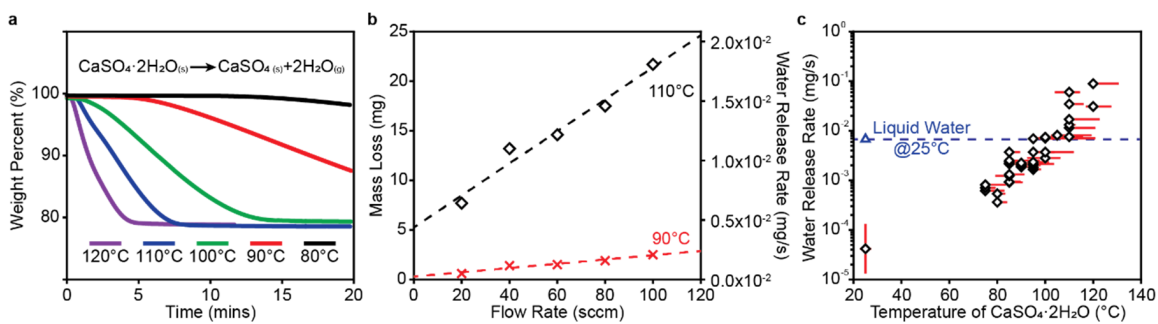

**Figure 1.** (a) Thermal decomposition of 10 mg $CaSO_4 \cdot 2H_2O$ at isothermal conditions as a function of time at different temperatures, showing definite water vapor release rate at different temperature, to eventually 79.1 wt% corresponding to the complete loss of water. (b) The experimental water vapor release rates of $CaSO_4 \cdot 2H_2O$ at different temperatures shows a linear dependence on the flow rate of carrier gas. The rate was measured by measuring the mass loss of ~1g $CaSO_4 \cdot 2H_2O$ before and after being held at a fixed temperature for 20 mins. (c) The relationship between the experimental water vapor release rate and the temperature of $CaSO_4 \cdot 2H_2O$, showing a wide dynamic range of controlled water vapor supply. All data points were measured with 100 sccm Ar flow and ~1 g initial mass of $CaSO_4 \cdot 2H_2O$. The horizontal error bars represent the range of the highest temperature measured before temperature was stabilized

and the lowest temperature during the reaction. The blue horizontal dash line represents the experimental water vapor release rate when liquid water in an alumina boat at 25 °C is used as the water source.

Although in a CVD system the decomposition rate could not be directly monitored *in operando*, the experimental water vapor release rate could be determined by measuring the mass change of $CaSO_4·2H_2O$ before and after reaction, as well as the time at the chosen elevated temperature. In a typical reaction where the temperature of $CaSO_4·2H_2O$ was controlled by heating tapes (See Experimental Section), only less than 3% of the total 1g mass of $CaSO_4·2H_2O$ would be released during the reaction time, which is much less than the 20.9% weight loss limit. The experimental water vapor release rate also shows a linear dependence on the flow rate of carrier gas at different temperatures (Figure 1b). This suggests that the released water vapor reaches an equilibrium with the $CaSO_4·2H_2O$ solid locally in the reactor before being swept away by carrier gas. A wide dynamic range of water vapor release rate can be achieved using $CaSO_4·2H_2O$ as the water source. As shown in Figure 1c, at a fixed Ar flow rate of 100 sccm, the experimental water vapor release rate can be tuned from $10^{-4}$ to $10^{-1}$ mg/s in the temperature range from 80 to 120 °C, which is equivalent to a flow rate of $10^{-3}$ to $10^0$ sccm of pure water vapor. In comparison, the water vapor release rate of 0.1 mL liquid water placed in an alumina boat was measured to be around $7 \times 10^{-3}$ mg/s under 100 sccm Ar flow (marked as the blue dashed line in Figure 1c). Although water vapor can also be delivered by using a gas bubbler, it requires a more complicated setup, and has more limited range of release rate, especially at the more critical lower range. In contrast, heating $CaSO_4·2H_2O$ solid provides a more controllable delivery of water vapor in a much wider range without using complicated setups.

We can utilize this controllable water vapor delivery to enable the reproducible vapor phase growth of MX$_2$. Although the growth method can be generally applied to WS$_2$, MoS$_2$, WSe$_2$, MoSe$_2$ (Figure S1), we are focusing on optimizing and understanding the growth of WS$_2$. In a typical growth reaction of WS$_2$ (Figure 2a, see Experimental Sections for details), CaSO$_4$·2H$_2$O was placed at the upstream far away from the three-zone tube furnace. This allows CaSO$_4$·2H$_2$O to stay at room temperature even when the furnace was heated to a high temperature, making it possible to use heating tapes to precisely control the temperature of CaSO$_4$·2H$_2$O during the reaction. The WS$_2$ precursor was placed in the center of Zone 1, and a large piece of 300 nm SiO$_2$/Si substrate was placed between Zone 2 and Zone 3, under an Argon flow of 100 sccm at 800 torr to minimize the penetration of ambient moisture into the reactor. When CaSO$_4$·2H$_2$O, Zone 2 and Zone 3 were heated up to the desired water precursor temperature (T$_w$, varied from 80 °C to 120 °C), 1200 °C and 800 °C (See a representative example of the actual temperature profiles in Figure S2), respectively and stabilized, the WS$_2$ precursor was pushed into Zone 2 to initiate the reaction. The water vapor released from CaSO$_4$·2H$_2$O is transported downstream by the carrier gas and reacts with WS$_2$ precursor at 1200 °C, which form volatile WO$_x$(OH)$_y$ gas species and H$_2$S gas,[35] as follows:

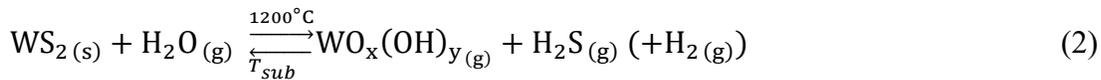

$$WS_{2\,(s)} + H_2O_{(g)} \underset{T_{sub}}{\overset{1200°C}{\rightleftarrows}} WO_x(OH)_{y\,(g)} + H_2S_{(g)}\;(+H_{2\,(g)}) \qquad (2)$$

These gaseous intermediates are transported downstream and react to form WS$_2$ on the substrate at lower temperature (Equation 2, the reverse direction). Note the equations are written in a general form because the oxidation states of tungsten might change during the reactions, which lead to the formation of hydrogen. The more specific reaction mechanisms still need further research. Essentially, when water vapor is present, the WS$_2$ growth reaction does not take place through a physical evaporation of WS$_2$ as commonly assumed, but rather water serves as the transport reagent for the chemical vapor transport of WS$_2$. Such CVT mechanism is supported by the observation of various tungsten oxide species formed on WS$_2$ precursor and substrate as the side products (Figure S3), particularly at high water vapor release rate conditions:

$$WO_x(OH)_{y\,(g)} \xrightarrow{T_{sub}} WO_{x\,(s)} + H_2O_{(g)} \qquad (3)$$

Although the CVT mechanism would generate WO$_x$(OH)$_y$ and H$_2$S stoichiometrically, due to the formation of tungsten oxide, WO$_x$(OH)$_y$ species in gas phase would be consumed faster than H$_2$S. As the reaction goes on, the ratio of H$_2$S and WO$_x$(OH)$_y$ could increase to slightly larger than the stochiometric ratio of 2:1 at different positions, which leads to the triangular and truncated triangular shape of MX$_2$ products, as previously reported.[57-58]

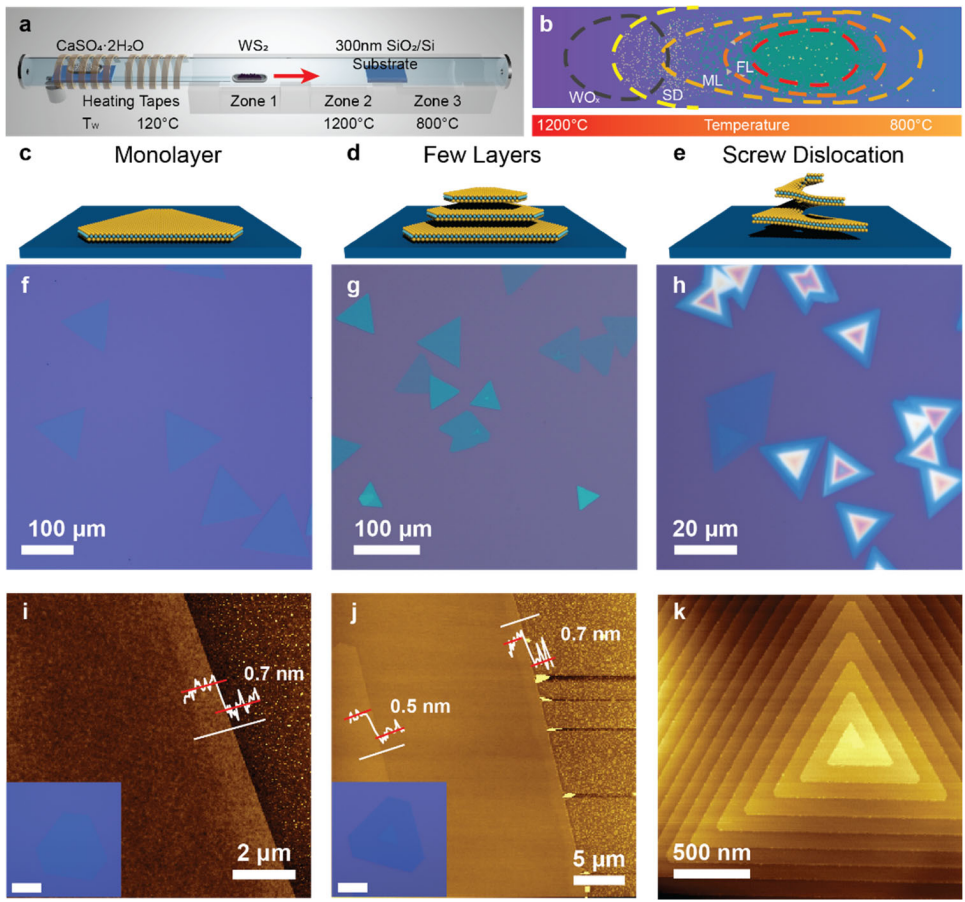

**Figure 2.** (a) Schematic illustration of the experimental setup of $WS_2$ growth by water-assisted chemical vapor transport. The $WS_2$ precursor was pushed into Zone 2 after all temperatures were stable. (b) Schematic illustration of the representative spatial distribution of different $WS_2$ products on the $SiO_2$/Si substrate: spiral nanoplates (SD, yellow), monolayer (ML, amber color), and few layers (FL, orange). The region enclosed by black dashed circle is the major growth region of $WO_x$ by product. The schematic structures (c-e), the optical (f-h) and atomic force microscopy images (i-k) of various $WS_2$ products: (c, f, i) monolayer, (d, g, j) few-layer nanoplate, (e, h, k) nanoplates with screw dislocations. The scale car in insets of (i, j) are 50 μm.

We consistently found different $WS_2$ products on different regions of the substrate (Figure 2b) when the $CaSO_4 \cdot 2H_2O$ was heated (i.e. the $T_w$) between 100 °C and 110 °C. Monolayer (ML, Figure 2c), few layers (FL, Figure 2d), and thick spiral nanoplates grown via a screw dislocation driven mechanisms (SD, Figure 2e) are the three typical and well-defined morphologies of single crystalline structures, as shown in optical microscope images (Figure 2f-h), as well as atomic force microscope (AFM) images and height profiles (Figure 2i-k). The lateral size of monolayer and few layer $WS_2$ plates are typically around 100 μm, with the thickness of few layer plates varying from 2 to 5 layers. Representative larger area optical images of the various types of products are shown in Figure S4. These morphologies are spatially distributed along the substrate in a predictable sequence. For example, when the $CaSO_4 \cdot 2H_2O$ was heated to 100 °C, along the gas flow direction from the temperature zone of ~1100 °C to ~900 °C on the substrate, one can find morphologies following a sequence of spiral plates, monolayer $WS_2$, and few layers $WS_2$. From ~900 °C to ~800 °C, the reverse sequence happens (Figure 2b). Besides such spatial distribution of morphologies, the more advantageous aspect of this water vapor assisted CVT method is that we now have the ability to control the morphologies in a specific region of the substrate with controlled delivery of water vapor by changing the heating temperature of $CaSO_4 \cdot 2H_2O$ ($T_w$). Specifically, by heating $CaSO_4 \cdot 2H_2O$ to ~85 °C, monolayer and dislocated $WS_2$ are the dominant morphologies at the temperature zone of around 1000 °C on the substrate. Further increasing $T_w$ to 120 °C would sequentially yield more few layer growth and polycrystalline growth but less monolayer growth in the same substrate region; while decreasing $T_w$ would sequentially yield more screw dislocation growth when $T_w$ is ~80 °C. Eventually, when $T_w$ is below 75 °C, there was no growth on the substrate (an example is shown in Figure S4a). The typical spatial distributions of various growth products under different $T_w$ conditions are summarized in Table S1 in the Supporting Information.

These crystal growth behaviors can be explained by the Burton-Cabrera-Frank (BCF) theory of crystal growth.[59-60] The crystal growth process is controlled by the supersaturation ($\sigma$) of local growth environment, which is defined as $\sigma = \ln(c/c_0)$, where $c$ and $c_0$ are the local concentration of gas precursor and equilibrium concentration of gas precursor, respectively. In the vapor phase growth of $MX_2$, low supersaturation promotes screw dislocation-driven growth mode, which results in spiral nanoplates, whereas high supersaturation leads to layer-by-layer growth mode characterized by monolayers as well as few layer nanoplates.[40, 45, 61] Here in our experiments, the local concentration $c$ is influenced by three factors: the equilibria of the reactions between $WS_2$ precursor and water vapor at 1200 °C that control the generation of gas precursors; the transport and distribution of gaseous species along the tube reactor; as well as the consumption of gas precursor during the deposition before reaching a particular growth region. Although the latter two factors are hard to predict and control, we can actively control the generation of gas precursors by tuning the amount of water vapor: higher vapor pressure of water in the reactor leads to higher local concentration $c$ for $WS_2$ growth. On the other hand, the equilibrium concentration $c_0$ is determined by the thermodynamic equilibrium of the $WS_2$ deposition reaction, which is only affected by the local temperature. If local concentration c is fixed, as temperature decreases along the tube reactor, the equilibrium concentration $c_0$ also decreases, which results in increasing supersaturation. Therefore, on the upstream side of the substrate where the local concentration c has not been consumed much, the supersaturation increases along the substrate; however, when the gas precursors are progressively consumed downstream, the supersaturation would decrease. This variation of supersaturation leads to the observed spatial distributions of various $WS_2$ morphologies along the substrate (Figure 2b) at a particular water vapor release rate.

Furthermore, the ability to control supersaturation also enables the controlled edge epitaxial growth of lateral $MX_2$ heterostructures. The key is to control the delivery of water vapor properly for each $MX_2$ material. We successfully grew large-area $WS_2$-$MoS_2$ heterostructures by switching the $MX_2$ source during the growth. As shown in Figure 3a, while $WS_2$ was reacting at 1200 °C in Zone 2, a second precursor boat containing $MoS_2$ placed upstream would stay cool and unreactive in Zone 1, where the temperature is around 500 °C during the reaction (Figure S5). After the $WS_2$ growth is completed, the $MoS_2$ precursor was pushed into Zone 2 as the $WS_2$ precursor was pushed out of Zone 2 using magnet coupled positioner and quartz rods. This allowed the *in situ* switch from $WS_2$ growth to $MoS_2$ growth without exposure to ambient conditions. One key to the success of lateral heterostructure growth is that the water source temperature $T_w$ needs to be lowered to ~85 °C, which leads to a low supersaturation of $MoS_2$ that only allows the $MoS_2$ growth seeded around the edges of the existing $WS_2$ layers via lateral epitaxy instead of self-nucleation. With this lateral epitaxy, three common types of heterostructures can be achieved as determined by the morphology of the $WS_2$ products in the first step, as illustrated in Figure 3b-d: lateral epitaxy on $WS_2$ monolayers would produce typical lateral heterostructures of monolayers (Figure 3b), however, lateral epitaxy on few layers could result in two types of structures, depending on the morphology of the multilayer WS$_2$ in the first step. In Type A where the layers of WS$_2$ have different sizes, lateral epitaxy would result in both lateral heterostructure and vertical heterostructure at different regions (Figure 3c); in Type B both layers have the same size and shape (Figure 3d), so only few layer lateral heterostructures would be produced. Because the starting few layer WS$_2$ plates are often found together, Type A and B are usually produced together on the substrate in high density (as shown in low magnification optical images in Figure S6).

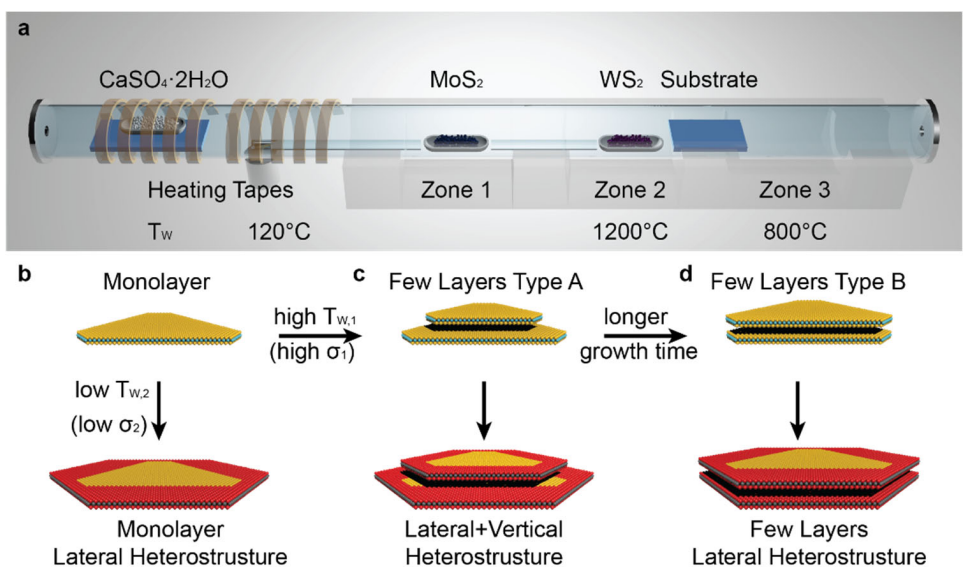

**Figure 3.** (a) Schematic illustration of the experimental setup for WS$_2$-MoS$_2$ heterostructure growth by water-assisted chemical vapor transport. The MoS$_2$ precursor was placed upstream of WS$_2$ precursor. When MoS$_2$ precursor was pushed into Zone 2, WS$_2$ precursor would simultaneously be pushed into Zone 3, which is downstream of the substrate. (b-d) The growth pathways of WS$_2$ and WS$_2$-MoS$_2$ heterostructures via lateral epitaxy leading to three common types of heterostructures.

We characterized these representative heterostructures using spatially resolved Raman and PL spectroscopy (Figure 4). These are large area heterostructures with lateral dimension greater than 100 μm. Figure 4a shows the optical microscope image of a monolayer $WS_2$-$MoS_2$ lateral heterostructure, where an optical contrast can be directly observed. The brighter region corresponds to the $WS_2$ grown in the first step, and the darker region corresponds to the $MoS_2$ grown in the second step. The PL spectra in Figure 4f (excited by 532 nm laser) show a strong peak at around 640 nm (1.94 eV) for $WS_2$ and 672 nm (1.84 eV) for $MoS_2$, which are consistant with previous reports.[35] The corresponding integrated PL intensity composite map of monolayer heterostrutures in Figure 4c shows good agreement with the optical contrast. Interestingly, spatial inhomogeneity was observed within each region. Three-fold "radioactive symbol" like pattern was observed in the $WS_2$ region, whereas the edges of $MoS_2$ region are significantly brighter, especially at the corners. In the corresponding Raman mapping (Figure 4b), where $WS_2$ was mapped by 2LA peak (531 cm$^{-1}$) and $MoS_2$ by $A_{1g}$ peak (406 cm$^{-1}$), the pattern in $WS_2$ region remained the same, but no obvious pattern was seen in $MoS_2$ region, suggesting different origins of inhomogeneity. The three-fold "radioactive symbol" like patterns are often observed in CVD synthesized $WS_2$, which was attributed to the spatial ditribution of different types and densities of point defects.[62-65] The enhanced PL on the edge of $MoS_2$ might be related to the oxygen binding to defects on the edge after the sample is exposed to ambient environment.[66-69] Note the boundaries between $WS_2$ and $MoS_2$ also showed brighter response in both PL and Raman. This could be related to potential charge transfer at the heterojunction of monolayer $WS_2$ and $MoS_2$,[70] which needs to be further invesigated.

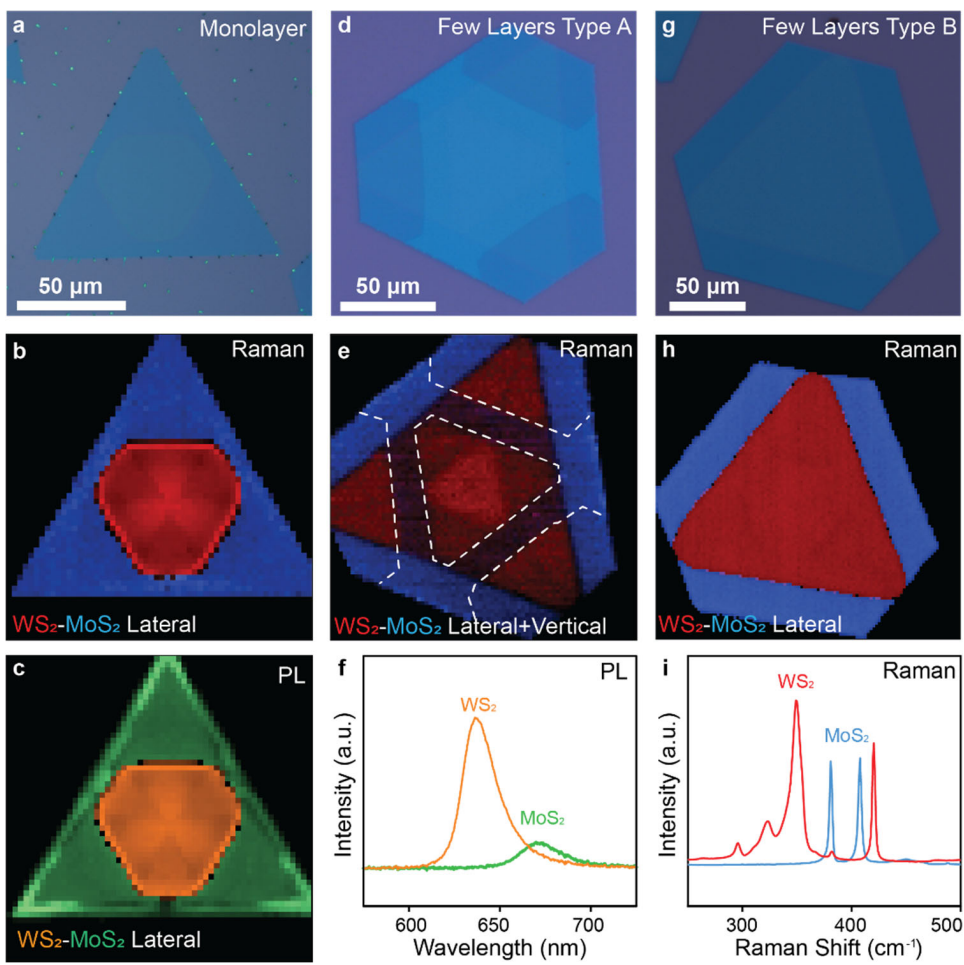

**Figure 4.** Optical image (a), Raman (b) and PL (c) mapping of a representative monolayer $WS_2$-$MoS_2$ heterostructure. Optical image (d) and Raman mapping (e) of a representative few layer type A $WS_2$-$MoS_2$ heterostructures. (f) PL Spectrum of the monolayer $WS_2$-$MoS_2$ heterostructure. Optical image (g) and Raman mapping (h) of a representative few layer type B $WS_2$-$MoS_2$ heterostructure. (i) Raman Spectrum of the few layer $WS_2$ -$MoS_2$ heterostructure. The samples were excited by 532 nm laser.

Raman mappings on the representative few layers type A and type B heterostructures were also performed. The particular type A heterostructures in Figure 4e was mapped by using 422 nm laser excitation instead of 532 nm, since 532 nm laser would enhance the Raman intensity of $WS_2$ via a double-resonant process[71] that buries the Raman signature of $MoS_2$ in the overlapped region. Non-resonant 442 nm laser would bring the peaks close to about equivalent intensity, thus helping to distinguish peaks from each material. The particular nanoplate in Figure 4b and 4e consists of a 3-layer base, which was confirmed by AFM as shown in Figure S7. In addition to that, another layer of smaller size $WS_2$ was grown during the first growth step. After the lateral epitaxial growth of $MoS_2$ in the second step, part of the extended $MoS_2$ region on the top layer overlaps with the $WS_2$ beneath, resulting in vertical heterostructures of $WS_2$-$MoS_2$. The overlapped region, as outlined by the dashed lines in Figure 4e, showed depressed Raman signals from both $WS_2$ and $MoS_2$ in comparison to the homostructure near it. The particular type B heterostructure in Figure 4g and 4h also consists of 3 layers of $WS_2$ and $MoS_2$, which was confirmed by AFM as shown in Figure S7. This few layer $WS_2$-$MoS_2$ lateral heterostructure showed the same difference in optical contrast seen in monolayer heterostructures, however the three-fold "radioactive symbol" is much less obvious. These Raman mapping results confirmed the various growth pathways and the type of $MX_2$ heterostructures that can be formed with water-assisted CVT (Figure 3).

In summary, we have demonstrated controllable synthesis of various $WS_2$ nanostructures and $WS_2$-$MoS_2$ heterostructures using water-assisted chemical vapor transport. By utilizing the thermal decomposition of $CaSO_4 \cdot 2H_2O$, we developed a simple method to precisely control the amount of water vapor in a CVD reactor. This method can not only provide the ability to generate water vapor with a wider range than using liquid water source, but also can be easily integrated into a CVD reaction setup. Controlled amount of water vapor, and thus controlled supersaturation of the CVT reactions, enabled us to achieve controllable growth of various $WS_2$ monolayer, few-layer and spiral nanoplates on $SiO_2$/Si substrates, the lateral epitaxial growth of $MoS_2$ layer on the edge of WS$_2$, and more reproducible growth of different types of large area heterostructures of WS$_2$-MoS$_2$. PL and Raman mappings confirmed three different types of heterostructures. This research provides a facile and effective approach to control the water vapor in vapor phase growth of MX$_2$ materials and reveal new insights into the growth mechanisms of MX$_2$, and thus leads to a general approach to the controllable growth of other metal chalcogenides and their heterostructures.

## ASSOCIATED CONTENT

**Supporting Information**.

Experimental sections; the representative optical images of MX$_2$ grown at different conditions; examples of the temperature profiles of the reactor system; powder X-ray diffraction of the WS$_2$ and MoS$_2$ precursors after reaction; summary of the morphology distribution for WS$_2$ growth at different T$_w$; optical images of WS$_2$ and WS$_2$-MoS$_2$ heterostructures. atomic force microscopy images WS$_2$-MoS$_2$ heterostructures.

## AUTHOR INFORMATION

**Corresponding Author**

*Correspondence should be addressed to S.J. (jin@chem.wisc.edu).

**Notes**

The authors declare no competing financial interest.

## ACKNOWLEDGMENT

This research is supported by the Department of Energy, Office of Basic Energy Sciences, Division of Materials Science and Engineering, under Award DE-FG02-09ER46664

REFERENCES

(1) Zeng, M.; Xiao, Y.; Liu, J.; Yang, K.; Fu, L. Exploring Two-Dimensional Materials toward the Next-Generation Circuits: From Monomer Design to Assembly Control *Chem. Rev.* **2018,** *118*, 6236.
(2) Novoselov, K. S.; Mishchenko, A.; Carvalho, A.; Castro Neto, A. H. 2D materials and van der Waals heterostructures *Science* **2016,** *353*, aac9439.
(3) Zhou, J.; Lin, J.; Huang, X.; Zhou, Y.; Chen, Y.; Xia, J.; Wang, H.; Xie, Y.; Yu, H.; Lei, J*., et al.* A library of atomically thin metal chalcogenides *Nature* **2018,** *556*, 355.
(4) Briggs, N.; Subramanian, S.; Lin, Z.; Li, X. F.; Zhang, X. T.; Zhang, K. H.; Xiao, K.; Geohegan, D.; Wallace, R.; Chen, L. Q*., et al.* A roadmap for electronic grade 2D materials *2D Mater.* **2019,** *6*, 022001.
(5) Ma, Y.; Ajayan, P. M.; Yang, S.; Gong, Y. Recent Advances in Synthesis and Applications of 2D Junctions *Small* **2018,** *14*, 1801606.
(6) Liu, B.; Abbas, A.; Zhou, C. Two-Dimensional Semiconductors: From Materials Preparation to Electronic Applications *Adv. Electron. Mater.* **2017,** *3*, 1700045.
(7) Liu, Y.; Weiss, N. O.; Duan, X.; Cheng, H.-C.; Huang, Y.; Duan, X. Van der Waals heterostructures and devices *Nat. Rev. Mater.* **2016,** *1*, 16042.
(8) Wang, Q. H.; Kalantar-Zadeh, K.; Kis, A.; Coleman, J. N.; Strano, M. S. Electronics and optoelectronics of two-dimensional transition metal dichalcogenides *Nat. Nanotechnol.* **2012,** *7*, 699.
(9) Radisavljevic, B.; Radenovic, A.; Brivio, J.; Giacometti, V.; Kis, A. Single-layer $MoS_2$ transistors *Nat. Nanotechnol.* **2011,** *6*, 147.
(10) Mak, K. F.; Lee, C.; Hone, J.; Shan, J.; Heinz, T. F. Atomically thin $MoS_2$: a new direct-gap semiconductor *Phys. Rev. Lett.* **2010,** *105*, 136805.
(11) Novoselov, K. S.; Jiang, D.; Schedin, F.; Booth, T. J.; Khotkevich, V. V.; Morozov, S. V.; Geim, A. K. Two-dimensional atomic crystals *Proc. Natl. Acad. Sci. U.S.A.* **2005,** *102*, 10451.
(12) Nicolosi, V.; Chhowalla, M.; Kanatzidis, M. G.; Strano, M. S.; Coleman, J. N. Liquid Exfoliation of Layered Materials *Science* **2013,** *340*, 1226419.
(13) Coleman, J. N.; Lotya, M.; O'Neill, A.; Bergin, S. D.; King, P. J.; Khan, U.; Young, K.; Gaucher, A.; De, S.; Smith, R. J*., et al.* Two-Dimensional Nanosheets Produced by Liquid Exfoliation of Layered Materials *Science* **2011,** *331*, 568.
(14) Manzeli, S.; Ovchinnikov, D.; Pasquier, D.; Yazyev, O. V.; Kis, A. 2D transition metal dichalcogenides *Nat. Rev. Mater.* **2017,** *2*, 17033.
(15) Chen, P.; Zhang, Z.; Duan, X.; Duan, X. Chemical synthesis of two-dimensional atomic crystals, heterostructures and superlattices *Chem. Soc. Rev.* **2018,** *47*, 3129.
(16) Chen, J.; Zhao, X.; Tan, S. J. R.; Xu, H.; Wu, B.; Liu, B.; Fu, D.; Fu, W.; Geng, D.; Liu, Y*., et al.* Chemical Vapor Deposition of Large-Size Monolayer $MoSe_2$ Crystals on Molten Glass *J. Am. Chem. Soc.* **2017,** *139*, 1073.

(17) Gong, Y.; Ye, G.; Lei, S.; Shi, G.; He, Y.; Lin, J.; Zhang, X.; Vajtai, R.; Pantelides, S. T.; Zhou, W., et al. Synthesis of Millimeter-Scale Transition Metal Dichalcogenides Single Crystals *Adv. Funct. Mater.* **2016,** *26*, 2009.
(18) Huang, J.-K.; Pu, J.; Hsu, C.-L.; Chiu, M.-H.; Juang, Z.-Y.; Chang, Y.-H.; Chang, W.-H.; Iwasa, Y.; Takenobu, T.; Li, L.-J. Large-Area Synthesis of Highly Crystalline $WSe_2$ Monolayers and Device Applications *ACS Nano* **2014,** *8*, 923.
(19) Li, M.-Y.; Shi, Y.; Cheng, C.-C.; Lu, L.-S.; Lin, Y.-C.; Tang, H.-L.; Tsai, M.-L.; Chu, C.-W.; Wei, K.-H.; He, J.-H., *et al.* Epitaxial growth of a monolayer $WSe_2$-$MoS_2$ lateral p-n junction with an atomically sharp interface *Science* **2015,** *349*, 524.
(20) Duan, X.; Wang, C.; Shaw, J. C.; Cheng, R.; Chen, Y.; Li, H.; Wu, X.; Tang, Y.; Zhang, Q.; Pan, A., *et al.* Lateral epitaxial growth of two-dimensional layered semiconductor heterojunctions *Nat. Nanotechnol.* **2014,** *9*, 1024.
(21) Gong, Y.; Lei, S.; Ye, G.; Li, B.; He, Y.; Keyshar, K.; Zhang, X.; Wang, Q.; Lou, J.; Liu, Z., *et al.* Two-Step Growth of Two-Dimensional $WSe_2$/$MoSe_2$ Heterostructures *Nano. Lett.* **2015,** *15*, 6135.
(22) Gong, Y.; Lin, J.; Wang, X.; Shi, G.; Lei, S.; Lin, Z.; Zou, X.; Ye, G.; Vajtai, R.; Yakobson, B. I., *et al.* Vertical and in-plane heterostructures from $WS_2$/$MoS_2$ monolayers *Nat. Mater.* **2014,** *13*, 1135.
(23) Gong, Y.; Lin, Z.; Ye, G.; Shi, G.; Feng, S.; Lei, Y.; Elías, A. L.; Perea-Lopez, N.; Vajtai, R.; Terrones, H., *et al.* Tellurium-Assisted Low-Temperature Synthesis of $MoS_2$ and $WS_2$ Monolayers *ACS Nano* **2015,** *9*, 11658.
(24) Lin, Y.-C.; Zhang, W.; Huang, J.-K.; Liu, K.-K.; Lee, Y.-H.; Liang, C.-T.; Chu, C.-W.; Li, L.-J. Wafer-scale $MoS_2$ thin layers prepared by $MoO_3$ sulfurization *Nanoscale* **2012,** *4*, 6637.
(25) Najmaei, S.; Liu, Z.; Zhou, W.; Zou, X.; Shi, G.; Lei, S.; Yakobson, B. I.; Idrobo, J.-C.; Ajayan, P. M.; Lou, J. Vapour phase growth and grain boundary structure of molybdenum disulphide atomic layers *Nat. Mater.* **2013,** *12*, 754.
(26) Wang, X.; Gong, Y.; Shi, G.; Chow, W. L.; Keyshar, K.; Ye, G.; Vajtai, R.; Lou, J.; Liu, Z.; Ringe, E., *et al.* Chemical Vapor Deposition Growth of Crystalline Monolayer $MoSe_2$ *ACS Nano* **2014,** *8*, 5125.
(27) Zhou, J.; Tang, B.; Lin, J.; Lv, D.; Shi, J.; Sun, L.; Zeng, Q.; Niu, L.; Liu, F.; Wang, X., *et al.* Morphology Engineering in Monolayer $MoS_2$-$WS_2$ Lateral Heterostructures *Adv. Funct. Mater.* **2018,** *28*, 1801568.
(28) Samad, L.; Bladow, S. M.; Ding, Q.; Zhuo, J.; Jacobberger, R. M.; Arnold, M. S.; Jin, S. Layer-Controlled Chemical Vapor Deposition Growth of $MoS_2$ Vertical Heterostructures via van der Waals Epitaxy *ACS Nano* **2016,** *10*, 7039.
(29) Zhang, X.; Meng, F.; Christianson, J. R.; Arroyo-Torres, C.; Lukowski, M. A.; Liang, D.; Schmidt, J. R.; Jin, S. Vertical Heterostructures of Layered Metal Chalcogenides by van der Waals Epitaxy *Nano. Lett.* **2014,** *14*, 3047.
(30) Yu, Y.; Li, C.; Liu, Y.; Su, L.; Zhang, Y.; Cao, L. Controlled Scalable Synthesis of Uniform, High-Quality Monolayer and Few-layer $MoS_2$ Films *Sci. Rep.* **2013,** *3*, 1866.
(31) Xie, S.; Tu, L.; Han, Y.; Huang, L.; Kang, K.; Lao, K. U.; Poddar, P.; Park, C.; Muller, D. A.; DiStasio, R. A., *et al.* Coherent, atomically thin transition-metal dichalcogenide superlattices with engineered strain *Science* **2018,** *359*, 1131.
(32) Eichfeld, S. M.; Colon, V. O.; Nie, Y.; Cho, K.; Robinson, J. A. Controlling nucleation of monolayer $WSe_2$ during metal-organic chemical vapor deposition growth *2D Mater.* **2016,** *3*, 025015.

(33) Kang, K.; Xie, S.; Huang, L.; Han, Y.; Huang, P. Y.; Mak, K. F.; Kim, C.-J.; Muller, D.; Park, J. High-mobility three-atom-thick semiconducting films with wafer-scale homogeneity *Nature* **2015**, *520*, 656.
(34) Lin, Y.-C.; Jariwala, B.; Bersch, B. M.; Xu, K.; Nie, Y.; Wang, B.; Eichfeld, S. M.; Zhang, X.; Choudhury, T. H.; Pan, Y*., et al.* Realizing Large-Scale, Electronic-Grade Two-Dimensional Semiconductors *ACS Nano* **2018,** *12*, 965.
(35) Sahoo, P. K.; Memaran, S.; Xin, Y.; Balicas, L.; Gutierrez, H. R. One-pot growth of two-dimensional lateral heterostructures via sequential edge-epitaxy *Nature* **2018,** *553*, 63.
(36) Zhang, Z.; Chen, P.; Duan, X.; Zang, K.; Luo, J.; Duan, X. Robust epitaxial growth of two-dimensional heterostructures, multiheterostructures, and superlattices *Science* **2017**, *357*, 788.
(37) Zheng, B.; Ma, C.; Li, D.; Lan, J.; Zhang, Z.; Sun, X.; Zheng, W.; Yang, T.; Zhu, C.; Ouyang, G*., et al.* Band Alignment Engineering in Two-Dimensional Lateral Heterostructures *J. Am. Chem. Soc.* **2018**, *140*, 11193.
(38) Duan, X.; Wang, C.; Fan, Z.; Hao, G.; Kou, L.; Halim, U.; Li, H.; Wu, X.; Wang, Y.; Jiang, J*., et al.* Synthesis of $WS_{2x}Se_{2-2x}$ Alloy Nanosheets with Composition-Tunable Electronic Properties *Nano. Lett.* **2016,** *16*, 264.
(39) Fan, X.; Jiang, Y.; Zhuang, X.; Liu, H.; Xu, T.; Zheng, W.; Fan, P.; Li, H.; Wu, X.; Zhu, X*., et al.* Broken Symmetry Induced Strong Nonlinear Optical Effects in Spiral $WS_2$ Nanosheets *ACS Nano* **2017**, *11*, 4892.
(40) Fan, X.; Zhao, Y.; Zheng, W.; Li, H.; Wu, X.; Hu, X.; Zhang, X.; Zhu, X.; Zhang, Q.; Wang, X*., et al.* Controllable Growth and Formation Mechanisms of Dislocated $WS_2$ Spirals *Nano. Lett.* **2018,** *18*, 3885.
(41) Huang, C.; Wu, S.; Sanchez, A. M.; Peters, J. J. P.; Beanland, R.; Ross, J. S.; Rivera, P.; Yao, W.; Cobden, D. H.; Xu, X. Lateral heterojunctions within monolayer $MoSe_2$–$WSe_2$ semiconductors *Nat. Mater.* **2014**, *13*, 1096.
(42) Cai, Z.; Liu, B.; Zou, X.; Cheng, H.-M. Chemical Vapor Deposition Growth and Applications of Two-Dimensional Materials and Their Heterostructures *Chem. Rev.* **2018**, *118*, 6091.
(43) Liu, K.-K.; Zhang, W.; Lee, Y.-H.; Lin, Y.-C.; Chang, M.-T.; Su, C.-Y.; Chang, C.-S.; Li, H.; Shi, Y.; Zhang, H*., et al.* Growth of Large-Area and Highly Crystalline $MoS_2$ Thin Layers on Insulating Substrates *Nano. Lett.* **2012**, *12*, 1538.
(44) Zhang, J.; Yu, H.; Chen, W.; Tian, X.; Liu, D.; Cheng, M.; Xie, G.; Yang, W.; Yang, R.; Bai, X*., et al.* Scalable Growth of High-Quality Polycrystalline $MoS_2$ Monolayers on $SiO_2$ with Tunable Grain Sizes *ACS Nano* **2014**, *8*, 6024.
(45) Shearer, M. J.; Samad, L.; Zhang, Y.; Zhao, Y.; Puretzky, A.; Eliceiri, K. W.; Wright, J. C.; Hamers, R. J.; Jin, S. Complex and Noncentrosymmetric Stacking of Layered Metal Dichalcogenide Materials Created by Screw Dislocations *J. Am. Chem. Soc.* **2017**, *139*, 3496.
(46) Zhang, L.; Liu, K.; Wong, A. B.; Kim, J.; Hong, X.; Liu, C.; Cao, T.; Louie, S. G.; Wang, F.; Yang, P. Three-Dimensional Spirals of Atomic Layered $MoS_2$ *Nano. Lett.* **2014**, *14*, 6418.
(47) Ly, T. H.; Zhao, J.; Kim, H.; Han, G. H.; Nam, H.; Lee, Y. H. Vertically Conductive $MoS_2$ Spiral Pyramid *Adv. Mater.* **2016**, *28*, 7723.
(48) Chen, L.; Liu, B.; Abbas, A. N.; Ma, Y.; Fang, X.; Liu, Y.; Zhou, C. Screw-Dislocation-Driven Growth of Two-Dimensional Few-Layer and Pyramid-like $WSe_2$ by Sulfur-Assisted Chemical Vapor Deposition *ACS Nano* **2014**, *8*, 11543.
(49) Kastl, C.; Chen, C. T.; Kuykendall, T.; Shevitski, B.; Darlington, T. P.; Borys, N. J.; Krayev, A.; Schuck, P. J.; Aloni, S.; Schwartzberg, A. M. The important role of water in growth of monolayer transition metal dichalcogenides *2D Mater.* **2017**, *4*, 021024.

(50) Hata, K.; Futaba, D. N.; Mizuno, K.; Namai, T.; Yumura, M.; Iijima, S. Water-Assisted Highly Efficient Synthesis of Impurity-Free Single-Walled Carbon Nanotubes *Science* **2004,** *306*, 1362.
(51) Amama, P. B.; Pint, C. L.; McJilton, L.; Kim, S. M.; Stach, E. A.; Murray, P. T.; Hauge, R. H.; Maruyama, B. Role of Water in Super Growth of Single-Walled Carbon Nanotube Carpets *Nano. Lett.* **2009,** *9*, 44.
(52) Zhou, W.; Zhan, S.; Ding, L.; Liu, J. General Rules for Selective Growth of Enriched Semiconducting Single Walled Carbon Nanotubes with Water Vapor as in Situ Etchant *J. Am. Chem. Soc.* **2012,** *134*, 14019.
(53) Asif, M.; Tan, Y.; Pan, L.; Li, J.; Rashad, M.; Usman, M. Thickness Controlled Water Vapors Assisted Growth of Multilayer Graphene by Ambient Pressure Chemical Vapor Deposition *J. Phys. Chem. C* **2015,** *119*, 3079.
(54) Wang, L.; Wu, B.; Liu, H.; Huang, L.; Li, Y.; Guo, W.; Chen, X.; Peng, P.; Fu, L.; Yang, Y., *et al.* Water-assisted growth of large-sized single crystal hexagonal boron nitride grains *Mater. Chem. Front.* **2017,** *1*, 1836.
(55) Moh, G. H. High-temperature metal sulfide chemistry *Top. Curr. Chem.* **1978,** *76*, 107.
(56) Brainard, W. A. The thermal stability and friction of the disulfides, diselenides, and ditellurides of molybdenum and tungsten in vacuum ($10^{-9}$ to $10^{-6}$ torr) *NASA Technical Note* **1969,** *TN D-5141*, 1.
(57) Dong, J.; Zhang, L.; Ding, F. Kinetics of Graphene and 2D Materials Growth *Adv. Mater.* *0*, 1801583.
(58) Wang, S.; Rong, Y.; Fan, Y.; Pacios, M.; Bhaskaran, H.; He, K.; Warner, J. H. Shape Evolution of Monolayer $MoS_2$ Crystals Grown by Chemical Vapor Deposition *Chem. Mater.* **2014,** *26*, 6371.
(59) Burton, W. K.; Cabrera, N.; Frank, F. C. The Growth of Crystals and the Equilibrium Structure of Their Surfaces *Philos. Trans. R. Soc. Lond. A* **1951,** *243*, 299.
(60) Woodruff, D. P. How does your crystal grow? A commentary on Burton, Cabrera and Frank (1951) 'The growth of crystals and the equilibrium structure of their surfaces' *Philos. Trans. R. Soc. A* **2015,** *373*, 20140230.
(61) Meng, F.; Morin, S. A.; Forticaux, A.; Jin, S. Screw Dislocation Driven Growth of Nanomaterials *Acc. Chem. Res.* **2013,** *46*, 1616.
(62) Lin, Y.-C.; Li, S.; Komsa, H.-P.; Chang, L.-J.; Krasheninnikov, A. V.; Eda, G.; Suenaga, K. Revealing the Atomic Defects of $WS_2$ Governing Its Distinct Optical Emissions *Adv. Funct. Mater.* **2018,** *28*, 1704210.
(63) Sheng, Y.; Wang, X.; Fujisawa, K.; Ying, S.; Elias, A. L.; Lin, Z.; Xu, W.; Zhou, Y.; Korsunsky, A. M.; Bhaskaran, H., *et al.* Photoluminescence Segmentation within Individual Hexagonal Monolayer Tungsten Disulfide Domains Grown by Chemical Vapor Deposition *ACS Appl. Mater. Interfaces* **2017,** *9*, 15005.
(64) Liu, H.; Lu, J.; Ho, K.; Hu, Z.; Dang, Z.; Carvalho, A.; Tan, H. R.; Tok, E. S.; Sow, C. H. Fluorescence Concentric Triangles: A Case of Chemical Heterogeneity in $WS_2$ Atomic Monolayer *Nano. Lett.* **2016,** *16*, 5559.
(65) Rosenberger, M. R.; Chuang, H.-J.; McCreary, K. M.; Li, C. H.; Jonker, B. T. Electrical Characterization of Discrete Defects and Impact of Defect Density on Photoluminescence in Monolayer $WS_2$ *ACS Nano* **2018,** *12*, 1793.

(66) Gogoi, P. K.; Hu, Z.; Wang, Q.; Carvalho, A.; Schmidt, D.; Yin, X.; Chang, Y.-H.; Li, L.-J.; Sow, C. H.; Neto, A. H. C., *et al.* Oxygen Passivation Mediated Tunability of Trion and Excitons in MoS$_2$ *Phys. Rev. Lett.* **2017,** *119*, 077402.

(67) Gutiérrez, H. R.; Perea-López, N.; Elías, A. L.; Berkdemir, A.; Wang, B.; Lv, R.; López-Urías, F.; Crespi, V. H.; Terrones, H.; Terrones, M. Extraordinary Room-Temperature Photoluminescence in Triangular WS$_2$ Monolayers *Nano. Lett.* **2013,** *13*, 3447.

(68) Kotsakidis, J. C.; Zhang, Q.; Vazquez de Parga, A. L.; Currie, M.; Helmerson, K.; Gaskill, D. K.; Fuhrer, M. S. Oxidation of Monolayer WS2 in Ambient Is a Photoinduced Process *Nano. Lett.* **2019,** *19*, 5205.

(69) Hu, Z.; Avila, J.; Wang, X.; Leong, J. F.; Zhang, Q.; Liu, Y.; Asensio, M. C.; Lu, J.; Carvalho, A.; Sow, C. H., *et al.* The Role of Oxygen Atoms on Excitons at the Edges of Monolayer WS2 *Nano. Lett.* **2019,** *19*, 4641.

(70) Shearer, M. J.; Li, M.-Y.; Li, L.-J.; Jin, S.; Hamers, R. J. Nanoscale Surface Photovoltage Mapping of 2D Materials and Heterostructures by Illuminated Kelvin Probe Force Microscopy *J. Phys. Chem. C* **2018,** *122*, 13564.

(71) Berkdemir, A.; Gutierrez, H. R.; Botello-Mendez, A. R.; Perea-Lopez, N.; Elias, A. L.; Chia, C. I.; Wang, B.; Crespi, V. H.; Lopez-Urias, F.; Charlier, J. C., *et al.* Identification of individual and few layers of WS$_2$ using Raman Spectroscopy *Sci. Rep.* **2013,** *3*, 1755.

TOC Graphic:

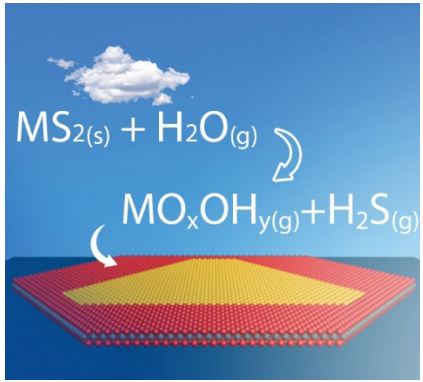